\documentclass[10pt,conference,letterpaper]{IEEEtran}
\PassOptionsToPackage{ruled, linesnumbered}{algorithm2e}
\usepackage[T1]{fontenc}
\usepackage[utf8]{inputenc}
\usepackage{cprotect}
\usepackage{float}
\usepackage{units}
\usepackage{algorithm2e}
\usepackage{graphicx}
\usepackage{xcolor}
\usepackage[caption=false,font=normalsize]{subfig}

\makeatletter

\usepackage{amsmath, amssymb, amsfonts, graphicx}

\allowdisplaybreaks

\makeatother

\begin{document}

\title{Accelerating $\mathrm{A}^*$-Based Algorithms for Decoding Quantum Low-Density Parity-Check Codes\\
}
\author{\IEEEauthorblockN{Lamia~Yous\IEEEauthorrefmark{1}, Francisco~Garcia Herrero\IEEEauthorrefmark{2}, and Mark~F.~Flanagan\IEEEauthorrefmark{1}}\IEEEauthorblockA{\IEEEauthorrefmark{1}School of Electrical and Electronic Engineering, University College Dublin, Belfield, Dublin 4, Ireland\\
\IEEEauthorrefmark{2}Department of Computer Architecture and Automatics, Complutense University of Madrid, Madrid, Spain\\
Email: lamia.yous@ucdconnect.ie, francg18@ucm.es, mark.flanagan@ieee.org }\vspace{-0.9cm} }


\maketitle

\begin{abstract}
Quantum low-density parity-check (QLDPC) codes represent a promising approach for error correction in quantum computing. The recently proposed Tesseract decoder uses the \(\text{A}^*\) search algorithm that guarantees finding the most likely error pattern \cite{b0}. However, practical implementations of Tesseract often involve an extremely large graph, and the inherently sequential nature of the search results in high computational overhead and long runtime. To improve decoding efficiency, we propose a two-stage decoding framework. First, a belief propagation (BP) decoder efficiently processes the syndrome. This step generates hard decisions (a binary error vector) and soft information (per-qubit confidence levels). In non-convergent BP cases, a gating mechanism examines the BP decoder's output to identify and filter out qubits with oscillating confidence values. In these cases, the refined output serves as input to Tesseract, which then attempts to identify and correct any residual errors. This hybrid approach leverages the high speed of BP decoding and delegates the more challenging decoding instances to Tesseract. Numerical results demonstrate a substantial reduction in overall decoding complexity while maintaining the logical error rate (LER) of the stand-alone Tesseract. Across all tested physical error rates, the proposed method achieves at least \(\mathbf{5\times}\) reduction in the number of expanded nodes in the Tesseract, with a peak reduction of nearly \(\mathbf{15\times}\) at a physical error rate of \(p=\mathbf{0.05}\) for the \(\mathrm{[[126,12, d<11]]}\) T1 code, and approximately \(\mathbf{8.8\times}\) at \(p = \mathbf{0.05}\) for the $[[72, 12, 6]]$ bicycle bivariate (BB) code.
\end{abstract}
\section{Introduction}
Quantum error correction (QEC) is crucial for enabling reliable, large-scale quantum computing by protecting fragile qubits from noise and decoherence \cite{b2}. To achieve fault-tolerant computing, logical qubits are encoded across multiple physical qubits \cite{b3}. A popular family of quantum codes is that of quantum low-density parity-check (QLDPC) codes; these are based on classical LDPC codes \cite{b4}, which are widely used in modern communication standards such as Wi-Fi and 5G \cite{b5}. QLDPC codes can achieve higher code rates and lower qubit overhead compared to surface codes.

The central practical issue for QLDPC codes is the design of an efficient decoder that can correct errors within the quantum state's strict decoherence time. If the decoding is too slow, the decoded error pattern may no longer accurately reflect the system state and cannot be used to efficiently rectify errors. Moreover, the rate at which syndromes are processed must match or exceed the rate at which syndromes are generated; otherwise, a small initial backlog in processing syndrome data will lead to an exponential slowdown during quantum computation \cite{b5.5}. Therefore, practical QEC requires both accurate and computationally efficient decoders.

One approach to decoding QLDPC codes is to employ message-passing decoding based on belief propagation (BP) algorithms, similar to those used for classical LDPC codes. BP offers a very fast solution for decoding; however, due to the presence cycles in the QLDPC Tanner graph, messages become correlated and may fail to converge. Furthermore, this algorithm exhibits a high error floor in some cases, meaning that even if the physical error rate in the quantum system improves, the logical error rate remains high \cite{b6} resulting in poor decoding performance. 

Google Quantum AI has recently published a search-based pathfinding decoder for QLDPC codes called Tesseract \cite{b0} which is based on the $\text{A}^*$ search algorithm. Tesseract guarantees to find the most likely error pattern. However, despite the use of pruning techniques, the decoding time grows rapidly with the size of the Tanner graph (i.e., the number of qubits $n$) and is still not efficient enough to be considered a real-time decoder. This limitation comes from the sequential nature of the $\text{A}^*$ search algorithm, which as a result cannot benefit from parallelism. More recent work has focused on optimizing the implementation of Tesseract through low-level performance enhancements \cite{b6.5}. However, these optimizations do not change the exponential search complexity and remain computationally expensive for large QLDPC codes.

Taken together, BP and Tesseract-based decoding approaches present a fundamental trade-off between computational efficiency and decoding performance, raising the question of whether their combination can achieve a superior trade-off. To address this challenge, we propose a new hybrid BP-Tesseract decoding framework that takes advantage of the complementary strengths of both approaches. In contrast to heuristic decision-tree decoders such as BP-DTD \cite{ott2025decision}, where BP is used to estimate costs for nodes in a decision tree (DT) which is then explored in a depth-first manner, our approach uses BP as a preprocessing stage to construct a full initial error estimate and associated reliability information.

The main contributions of the paper can be summarized below:
\begin{itemize}
    \item We propose a hybrid BP-Tesseract decoding framework in which the BP is first used to produce an initial error estimate from which a residual syndrome is computed, allowing Tesseract to operate on a reduced decoding problem. We further incorporate per-qubit reliability metrics derived from BP into the Tesseract search to guide and prioritize candidate error patterns.
    \item To ensure that only reliable BP soft information is propagated, we introduce a reliability-based gating mechanism that filters the BP confidence values before they are supplied to Tesseract.
    \item We validate our method using the T1 code from the T family of codes \cite{b7} and for the $[[72, 12, 6]]$ bicycle bivariate (BB) code \cite{bravyi2024high}, for X-type error correction.
\end{itemize}
\section{Preliminaries}
\subsection{QLDPC codes}
QLDPC codes are a class of QEC codes that extend the principle of classical sparse-graph codes to the quantum setting. A QLDPC code is defined such that the number of bits involved in each check and the number of checks acting on each bit are bounded by a constant, ensuring sparsity of the parity-check matrix \cite{b1}. Many QLDPC codes are constructed using the Calderbank-Shor-Steane (CSS) framework, i.e., they are defined by two classical binary linear codes $\mathbf{C}_X$ and $\mathbf{C}_Z$ \cite{b3}, \cite{7.75}. This construction yields two sparse binary parity-check matrices $\mathbf{H}_X$ and $\mathbf{H}_Z$ satisfying the orthogonality condition $\mathbf{H}_X\mathbf{H}_Z^T = 0$, ensuring that the X and Z-type errors commute. Due to the CSS structure, X-type and Z-type errors can be decoded independently. Since the decoding procedure is structurally identical for both types of error, we focus on decoding of X-type errors without loss of generality. 

A QLDPC code is typically represented using a Tanner graph. The graph includes check nodes, corresponding to the rows of the parity-check matrix $\mathbf{H}$, and variable nodes corresponding to its columns. The check nodes enforce parity-check constraints through their connections to the variable nodes. Decoding involves finding an error pattern that satisfies the parity-check equations defined by the measured syndrome. Let $\mathbf{e}\in \mathbb{F}_2^n$ denote the binary error vector and $\mathbf{s}\in \mathbb{F}_2^m$ the measured syndrome, where $n$ is the number of variable nodes and $m$ is the number of check nodes. Let $\mathbf{H}$ denote the parity-check matrix corresponding to the error type considered. The decoder attempts to find the most likely error pattern $\mathbf{e}$ that is consistent with the measured syndrome $\mathbf{s}$, i.e., that satisfies 
\begin{equation}
    \mathbf{s} = {\mathbf{H}}\mathbf{e} \pmod{2}\text{.}
    \label{syndrome_equation}
\end{equation}
where vector-matrix multiplication is over the binary field.

\subsection{Tesseract Decoder \cite{b0}}
Tesseract uses the $\text{A}^*$ search algorithm with an admissible heuristic and incorporates pruning strategies to improve speed. Under an independent noise model, each qubit error occurs with probability $p_i \in (0, 1/2]$. The decoding problem takes the form
\begin{equation}
    \hat{\mathbf{e}} = \arg\min_{\mathbf{e}\in\mathbb{F}_2^n} w(\mathbf{e}) \text{ subject to }\mathbf{H}\mathbf{e} = \mathbf{s},
\end{equation}
where the cost function $w(\mathbf{e})$ is defined as
\begin{equation}
    w(\mathbf{e})=\sum_{i=1}^n-e_i \log\left(\frac{p_i}{1-p_i}\right).
    \label{EQ: Cost function}
\end{equation}
The decoding task is to identify an error vector $\mathbf{e}$ that matches the measured syndrome (i.e., satisfies (\ref{syndrome_equation})) and minimizes the cost $w(\mathbf{e})$. 
In a graph-based approach to solving this optimization problem, the decoding task becomes a shortest-path search. Each node in the search graph represents a candidate error vector, and the initial node corresponds to the zero vector. An edge between two nodes corresponds to the addition of a single qubit error to the current estimate. The cost of traversing any edge equals the weight associated with the corresponding qubit error. The total cost of a path is the cost of the associated error vector. In the Tesseract graph, each node is additionally associated with a residual syndrome which is defined as
\begin{equation}
    \mathbf{s}_{\mathrm{res} }= \mathbf{s}\oplus \mathbf{H}\mathbf{e}'.
\end{equation}
The search begins from the zero error vector, and the goal is to incrementally add candidate qubit errors until the residual syndrome is equal to zero. At each node expansion, the algorithm selects the node with the minimum estimated total cost
\begin{equation}
    f(\mathbf{e}') = g(\mathbf{e}') + h(\mathbf{e}'),
\end{equation}
where $g(\mathbf{e}')$ is the accumulated cost of the current error pattern and $h(\mathbf{e}')$ is the heuristic function. A node expansion corresponds to selecting a candidate error vector and generating its neighboring candidates. In this paper, the total number of node expansions is used as a measure of decoding complexity.

The heuristic function is a lower bound on the additional cost required to correct the remaining syndrome, ensuring admissibility and guaranteeing optimality of the solution. It is calculated by checking the residual syndrome and estimating the minimum cost required to fix the remaining unsatisfied syndrome elements. To reduce the search space, Tesseract also applies pruning. The algorithm limits expansions to qubits connected to the lowest-indexed check nodes with a non-zero residual syndrome. Additional pruning rules prevent revisiting equivalent error configurations, which stops redundant search paths from being explored.

Although $\text{A}^*$ guarantees to find the most likely error and pruning strategies are employed to speed up the decoding process, in practical applications the graph involved can be extremely large, leading to a prohibitive time and complexity cost.
\subsection{Belief-Propagation Decoder \cite{b4}}
The BP decoder is an iterative message-passing decoder that operates on the code's Tanner graph in order to estimate the most likely binary error vector $\hat{\mathbf{e}}$ that matches the measured syndrome $\mathbf{s}$. Messages are exchanged along graph edges between variable and check nodes over multiple iterations. This section gives an overview; more details can be found in \cite{b8} and \cite{b8.5}.

Let $\boldsymbol{\mu} \in \mathbb{R}^n$ be the channel LLR vector, with i-th entry given by
\begin{equation}
    \mu_{i} = \log\frac{P(e_i = 0)}{P(e_i = 1)}.
\end{equation}
At the start ($t=0$), the variable-to-check messages are set using the channel LLRs for every edge in the Tanner graph, i.e.,
\begin{equation}
    m_{i\rightarrow j}^{(0)} = \mu_i.
\end{equation}
Here, $m_{i\rightarrow j}^{(t)}$ represents the message sent from variable node $i$ to check node $j$ in iteration $t$, while $m_{j\rightarrow i}^{(t)}$ is the message sent from check node $j$ to variable node $i$ in iteration t. $\mathcal{N}(i)$ is the set of check nodes connected to variable node $i$, and $\mathcal{N}(j)$ is the set of variable nodes connected to check node $j$. In each iteration, from $t=1$ to $T$, messages are updated along all edges. The messages from variable nodes to check nodes are updated via
\begin{equation}
    m_{i \rightarrow j}^{(t)} = \mu_{i} + \sum_{j'\in \mathcal{N}(i)\setminus j}m_{j'\rightarrow i}^{(t-1)},
\end{equation}
while messages from check nodes to variable nodes are updated via
\begin{equation}
    m_{j\rightarrow i}^{(t)} = (-1)^{s_{ j}}2\tanh^{-1}\prod_{i'\in \mathcal{N}(j)\setminus{i}} \tanh(m_{i'\rightarrow j}^{(t)}/2).
\end{equation}
After each iteration, the posterior log-likelihood ratio can be computed as 
\begin{equation}
    L_i^{(t)} = \mu_i+\sum_{j\in \mathcal{N}(i)}m_{j\rightarrow i}^{(t)}.
\end{equation}
After several iterations, a hard decision is formed from the accumulated LLRs as follows:
\begin{equation}
    \hat{e}_i = \begin{cases}
        0,\quad L_i^{(t)} > 0,\\
        1,\quad L_i^{(t)} \leq 0.
        
    \end{cases}
\end{equation}
Decoding is successful if the estimated error satisfies
\begin{equation}
    \mathbf{s} = \mathbf{H}\hat{\mathbf{e}}
\label{EQ:syndrome equation with e hat}
\end{equation} 
or if the maximum number of iterations $T_{\mathrm{max}}$ is reached. In QLDPC codes, Tanner graphs often have many short cycles, which can cause messages to become correlated. Because of this, the decoder may show oscillatory behavior when it does not converge \cite{b8.9}.

Despite these limitations, BP has computational complexity that is linear in the number of edges in the Tanner graph per iteration and can process nodes in parallel, rendering it a fast decoder. Rather than using the BP as a stand-alone decoder, we leverage its soft output to construct reliability-informed inputs that guide the Tesseract search. This enables a more informed search in non-convergent BP cases.
\section{Proposed Method}
\subsection{Hybrid Decoding Pipeline}
We assume an independent and identically distributed (i.i.d.) symmetric depolarizing error model, where each qubit undergoes a depolarizing error independently with probability $p$. The resulting syndrome measurements are first provided to a preliminary decoder, specifically, a BP decoder. The BP decoder's role is to attempt to identify the most likely error vector consistent with the measured syndrome. If BP successfully converges to a valid solution, its output is used as the final result, and no further steps are needed. If BP does not converge, further decoding is performed by Tesseract.

The BP decoder produces the LLRs $L_i^{t}$, $ t=1,2,\ldots,T_{\mathrm{max}}$, which can be converted to error probabilities in the linear domain. Let
\begin{equation}
    p_{\mathrm{err},i}^{(t)} = \frac{1}{1+\exp\left(L_i^{(t)}\right)}
\label{EQ:LLR to p conversion}
\end{equation}
denote the BP-estimated error probability of qubit $i$ at iteration $t$. The final BP output is then given by
\begin{equation}
    p_{\mathrm{BP}, i} = p_{\mathrm{err}, i}^{(T_{\mathbf{max}})}
\label{EQ:PBP}
\end{equation}
where $T_{\mathrm{max}}$ is the final BP iteration. Values of $p_{\mathrm{BP}, i}>0.5$ indicate that qubit $i$ is more likely to contain an error, while values below 0.5 indicate that it is more likely to contain no error.

The soft outputs of the BP decoder, which are used as input to Tesseract, provide an estimate of the probability that each BP hard decision is incorrect. These values are
\begin{equation}
    p_{\mathrm{flip}, i} = \min\left(p_{\mathrm{BP}, i},\, 1 - p_{\mathrm{BP}, i}\right), \quad i = 1,\dots,n.
\label{pflipequation}
\end{equation}
The value of $p_{\mathrm{flip}}$ provides a measure of uncertainty in the BP hard decision. Highly confident decisions, i.e., ${p_{\mathrm{BP}}}$ close to 0 or 1, should yield small values of $p_{\mathrm{flip}}$, while values near 0.5 relate to uncertain qubits. Essentially, this approach penalizes uncertain qubits and prioritizes those with low confidence during the search.

In the non-convergent case, the hard decisions from the BP and their corresponding $p_{\mathrm{flip}}$ values are extracted. The BP hard decisions are first used to construct an estimated error vector $\mathbf{e}_{\mathrm{BP}}$, where $\mathbf{e}_{\mathrm{BP},i}=0$ if $L_i>0$ and $\mathbf{e}_{\mathrm{BP},i}=1$ otherwise. From this, the corresponding syndrome is computed as $\mathbf{s}_{\mathrm{BP}} = \mathbf{H}\mathbf{e}_{\mathrm{BP}}$. As part of the decoding framework, Tesseract is not applied directly to the measured syndrome. Instead, we compute a residual syndrome. This residual syndrome represents the discrepancy between the BP prediction and the measured syndrome, and is defined as
\begin{equation}
    \mathbf{s}_{\mathrm{res}} = \mathbf{s} \oplus\mathbf{s}_{\mathrm{BP}},
    \label{modified_syndrome}
\end{equation}
where $\mathbf{s}_{\mathrm{BP}}$ denotes the syndrome computed from the hard decisions made by the BP. Tesseract is then used to identify the lowest-cost correction $\Delta\mathbf{e}$ that satisfies the residual syndrome. The final decoded error vector is determined as
\begin{equation}
    \mathbf{e}_{\text{final}} = \mathbf{e}_{\text{BP}} \oplus \Delta \mathbf{e}.
    \label{True final error vector equation}
\end{equation}
\subsection{Preprocessing Decoder}
\begin{figure*}[tb]

\subfloat[Convergent case.\label{convergent_BP}]{
\includegraphics[width=0.48\textwidth]{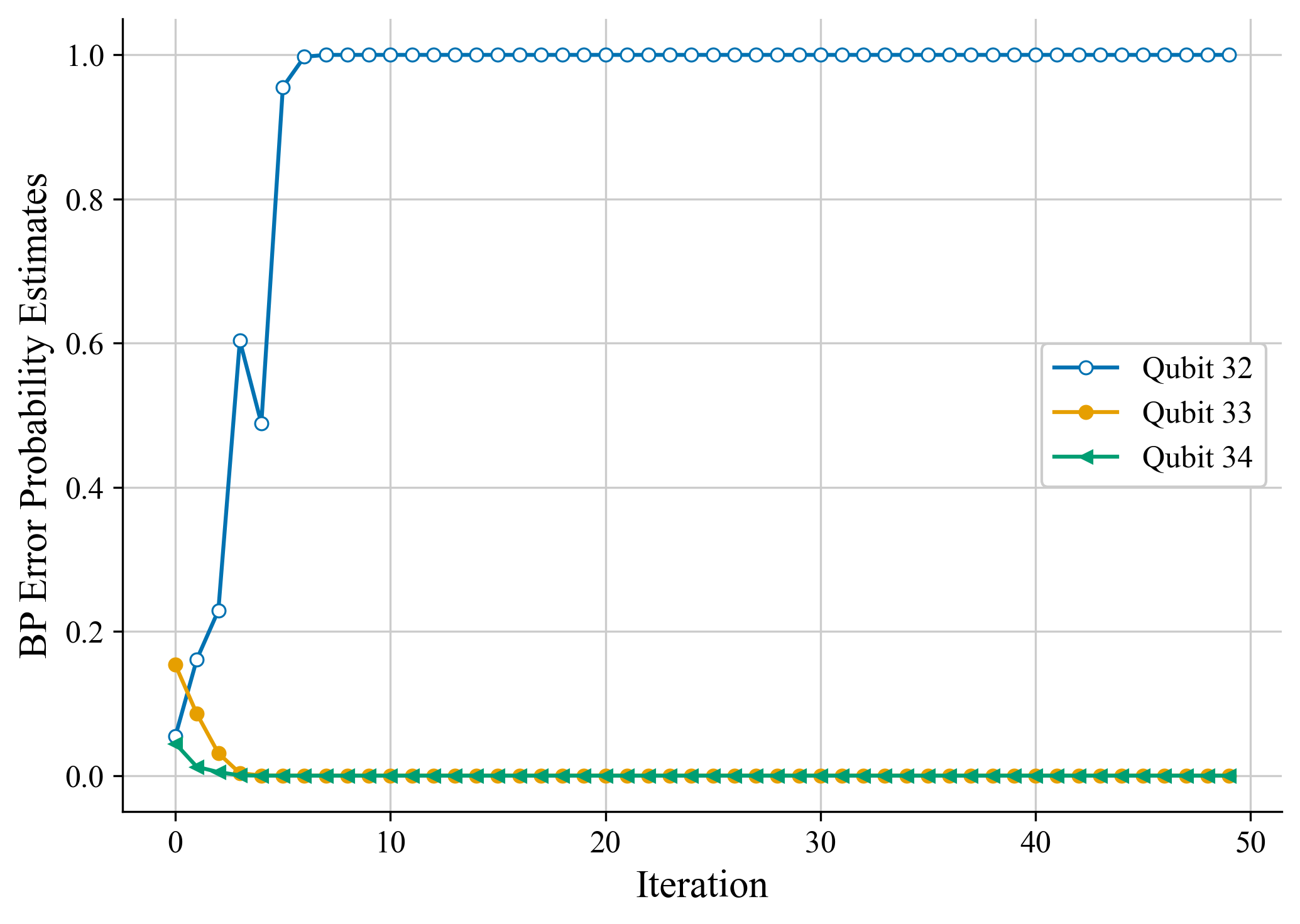}
}
\hfill
\subfloat[Non-convergent case.\label{non_convergent_BP}]{
\includegraphics[width=0.48\textwidth]{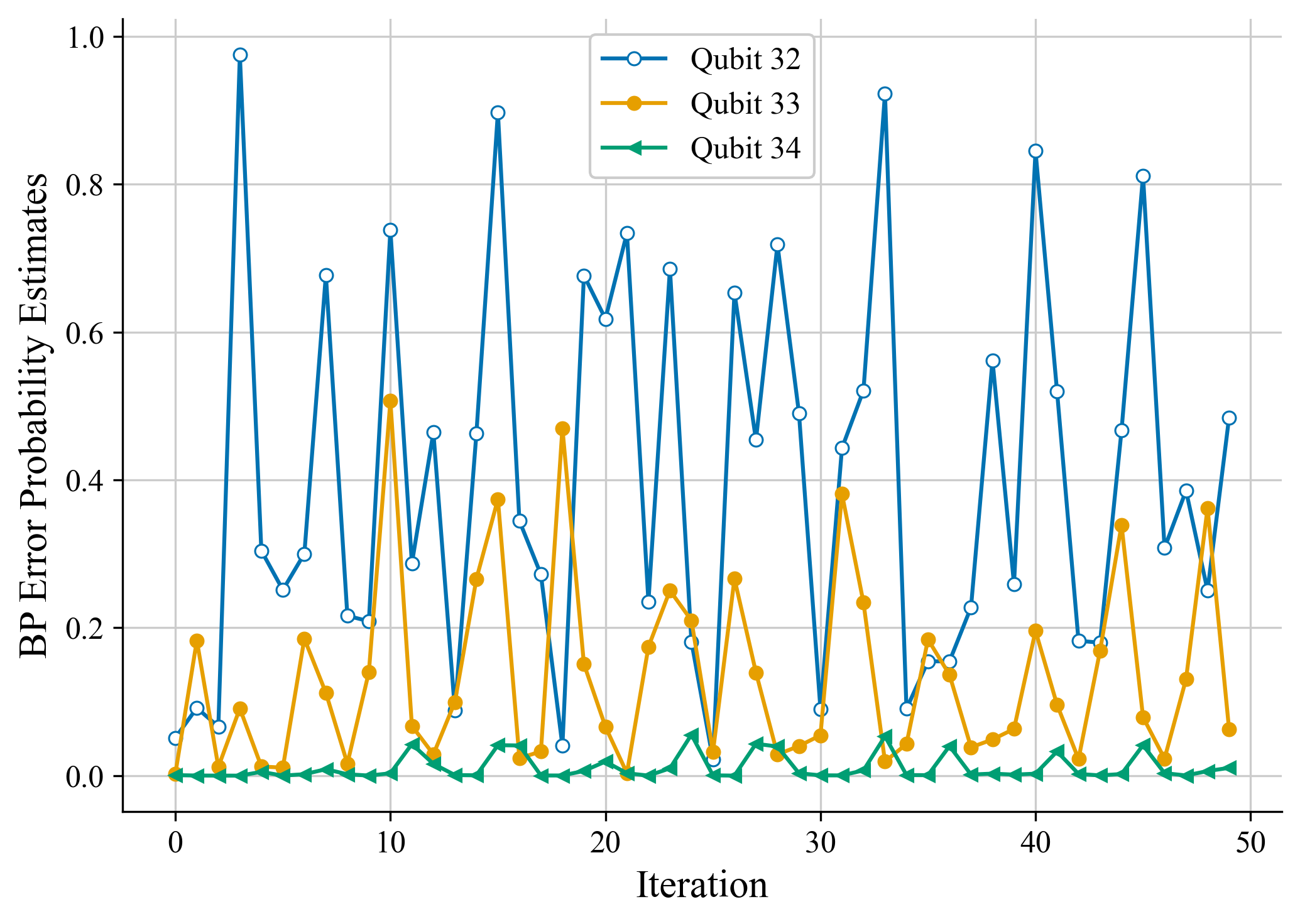}
}
\caption{Evolution of the estimated error probability from BP for a set of three sample qubits. The horizontal axis shows the BP iteration number, while the vertical axis shows the estimated probability of error. Each colored curve represents a different qubit in the Tanner graph. The curves are shown for qubits 32, 33 and 34 as these are representative of general qubit behavior.}
\label{convergentvsnon_convergentBP}
\end{figure*}
To reduce the overall decoding latency, BP decoding is used as a fast preprocessing step. When BP fails to converge, some qubits may exhibit oscillatory behavior in their LLRs. An example of this behavior is illustrated in Figure \ref{convergentvsnon_convergentBP} using the $[[126, 12, d<11]]$ T1 code, where the evolution of the error probabilities for qubits 32, 33 and 34 is shown as a function of BP iterations. In the convergent case shown in Figure \ref{convergent_BP}, the error probabilities for all three qubits stabilize after approximately 15 iterations, approaching either 0 (indicating no error) or 1 (indicating the presence of an error).

The non-convergent case illustrated in Figure \ref{non_convergent_BP} exhibits clear oscillatory behavior. For example, the error probability of qubit 32 fluctuates significantly from one iteration to the next. At iteration 45, the decoder estimates probability of error for this qubit to be greater than 80\%, whereas at iteration 48, this probability drops to approximately 25\%. This indicates that it repeatedly switches between predicting an error and no error. Qubit 33 also shows an oscillation, although its hard decisions remain unaltered. While the decoder consistently predicts no error for this qubit, its confidence values vary greatly, ranging from nearly 0\% error probability at iteration 46 to nearly 40\% at iteration 48. Although the hard decision is consistent, its error estimates are clearly unreliable. In contrast, qubit 34 shows converging behavior, with its error probability settling to a consistently low value.

\begin{figure}[tb]
    \centering
    \includegraphics[width=\columnwidth]{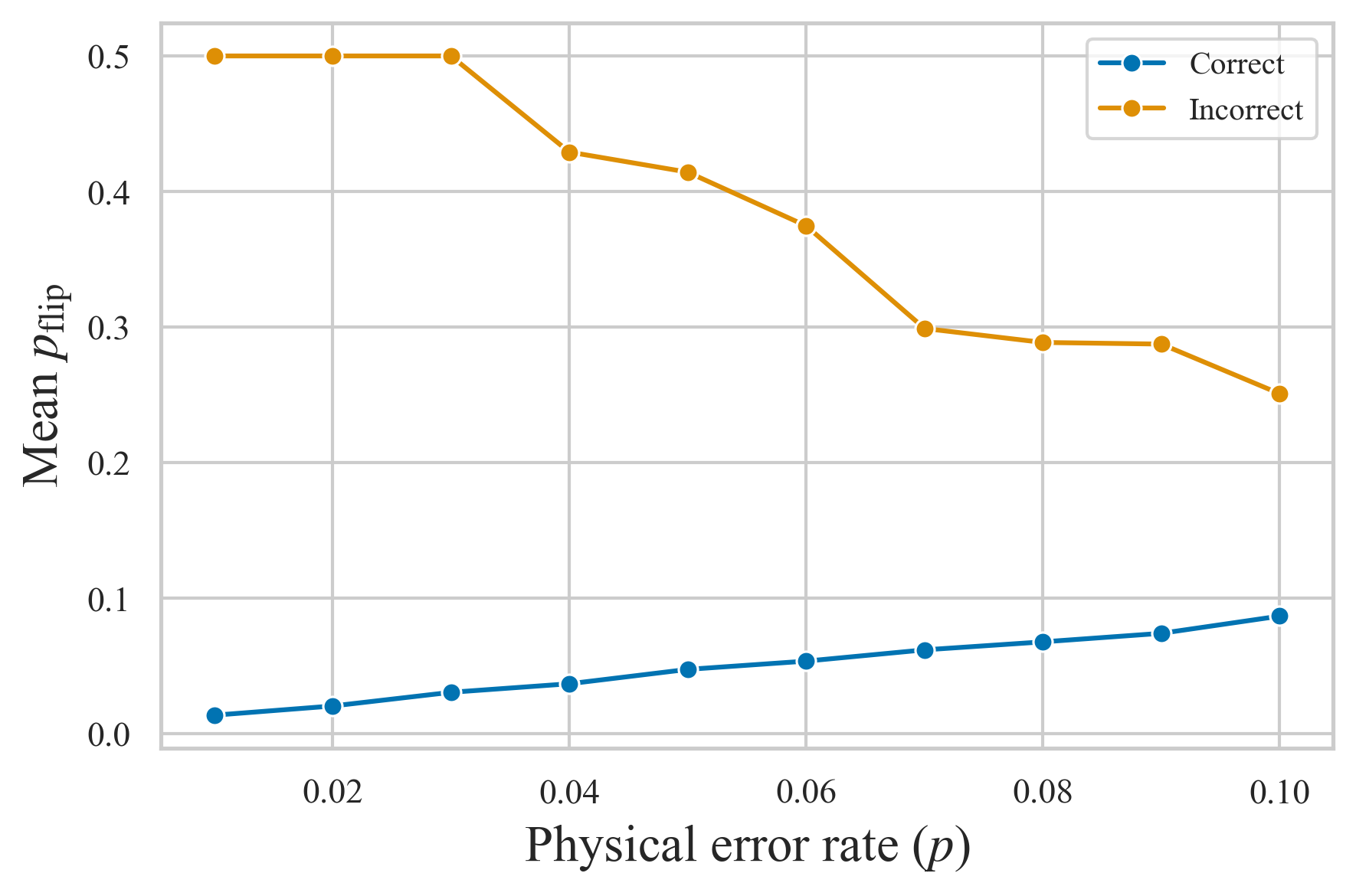}
    \caption{Mean $p_{\mathrm{flip}}$ values for correct and incorrect qubits versus physical error rate, $p$}
    \label{fig:correctmean}
\end{figure}

For BP to be an effective preprocessing stage, the soft information it provides must be reliable. Tesseract depends on this data to guide its search and can be misled by inaccurate estimates. To address this issue, we introduce a reliability-based gating mechanism that filters unreliable qubit estimates before passing the soft information to Tesseract. Only qubits whose confidence estimates are considered sufficiently reliable are allowed to influence the $\mathrm{A}^*$ search. In this work, reliability is determined using two main metrics. The first metric uses the value of ${p_{\mathrm{flip}}}$. To assess its suitability as a reliability measure, qubits are grouped according to whether the BP hard decision matches the true error pattern. Figure \ref{fig:correctmean} shows the mean value of ${p_{\mathrm{flip}}}$ per group as a function of the physical error rate $p$. It can be observed from the figure that correctly decoded qubits tend to have low ${p_{\mathrm{flip}}}$ values, indicating high BP confidence. This finding motivates our proposal of using a threshold to filter out qubits with large ${p_{\mathrm{flip}}}$ values.

However, the value of ${p_{\mathrm{flip}}}$ alone is not sufficient to determine the reliability. As shown by qubit 33 in Figure \ref{non_convergent_BP}, a qubit may exhibit a low error probability at the final iteration while still displaying significant oscillatory behavior during the decoding process. In such cases, the confidence value may appear reliable despite the estimate being oscillatory. Therefore, to quantify this oscillatory behavior, we introduce a second reliability metric that measures the temporal variation of the error probability across BP iterations. Oscillatory behavior is measured by computing the average step change in the probability of error across the final $20\%$ of BP iterations. We consider only the ability to detect persistent oscillations rather than transient fluctuations, which commonly occur in the early stages of BP decoding. As evidenced by the behavior of qubit 33 in Figure \ref{convergent_BP}, such early oscillations often stabilize as the decoder converges and should therefore not be penalized. The average step change is calculated as
\begin{equation}
    \Delta_{\mathrm{mean}, i} = \frac{1}{T-1}\sum_{t=T_{\max}-T}^{T_{\max}-1}\left|p_{\mathrm{err}, i}^{(t+1)}-p_{\mathrm{err}, i}^{(t)}\right|,\quad i = 1,\dots,n,
\label{step equation}
\end{equation}
where $T = T_{\max}/5$ and $p_{\mathrm{err},i}^{(t)}$ is the BP-estimated error probability of qubit $i$ at iteration $t$, as given in (\ref{EQ:LLR to p conversion}).

A qubit may have a very low ${p_{\mathrm{flip},i}}$, indicating high confidence in the BP hard decision; however, if its $\Delta_{\mathrm{mean},i}$ is large, the estimate is oscillatory and therefore unreliable. Conversely, qubits may exhibit a small $\Delta_{\mathrm{mean},i}$ but still have a high ${p}_{\mathrm{flip},i}$, indicating insufficient confidence in the BP estimate. In both cases, we decide not to use the BP-estimate to guide Tesseract. 

To simplify the optimization of these parameters, we define the overall reliability score for qubit i as $\alpha_{\mathrm{BP},i}=\Delta_{\mathrm{mean},i}{p}_{\mathrm{flip},i}$. This metric reflects both the confidence and temporal stability of each qubit estimate. A threshold $\tau$ is then used to classify each qubit as reliable or unreliable.

For qubits deemed to be unreliable, we assign a fixed prior probability of error $p_{\mathrm{gated}}$ for Tesseract, which represents a high level of uncertainty about the presence of an error on that qubit. On the other hand, for qubits deemed to be reliable, the original value $p_{\mathrm{flip}}$ is passed to Tesseract. The probabilities passed as priors to Tesseract are therefore
\begin{equation}
{p}_{\text{Tess},i} =
\begin{cases}
p_{\text{gated}}, & \text{if } \alpha_{\mathrm{BP},i}> \tau \\
{p}_{\text{flip},i}, & \text{otherwise}
\end{cases}\quad i = 1,\dots,n.
\end{equation}
The main task is to determine the optimal values for the threshold $\tau$, and $p_{\mathrm{gated}}$ values to significantly reduce the Tesseract search space. Algorithmic speed-up is defined by the reduction in the number of expanded nodes relative to Tesseract.
\subsection{Parameter Optimization}
In this section, we describe the method we have used to optimize $\tau$ and $p_{\mathrm{gated}}$, where we report results obtained with the T1 code as a representative example. The BP decoder from the LDPC decoding library developed by Joschka Roffe \cite{b9} was used. For all simulations, we used the sum-product decoder with serial scheduling and a maximum of $T_{\mathrm{max}}=50$ iterations.

\begin{figure}[tb]
    \centering
    \includegraphics[width=\columnwidth]{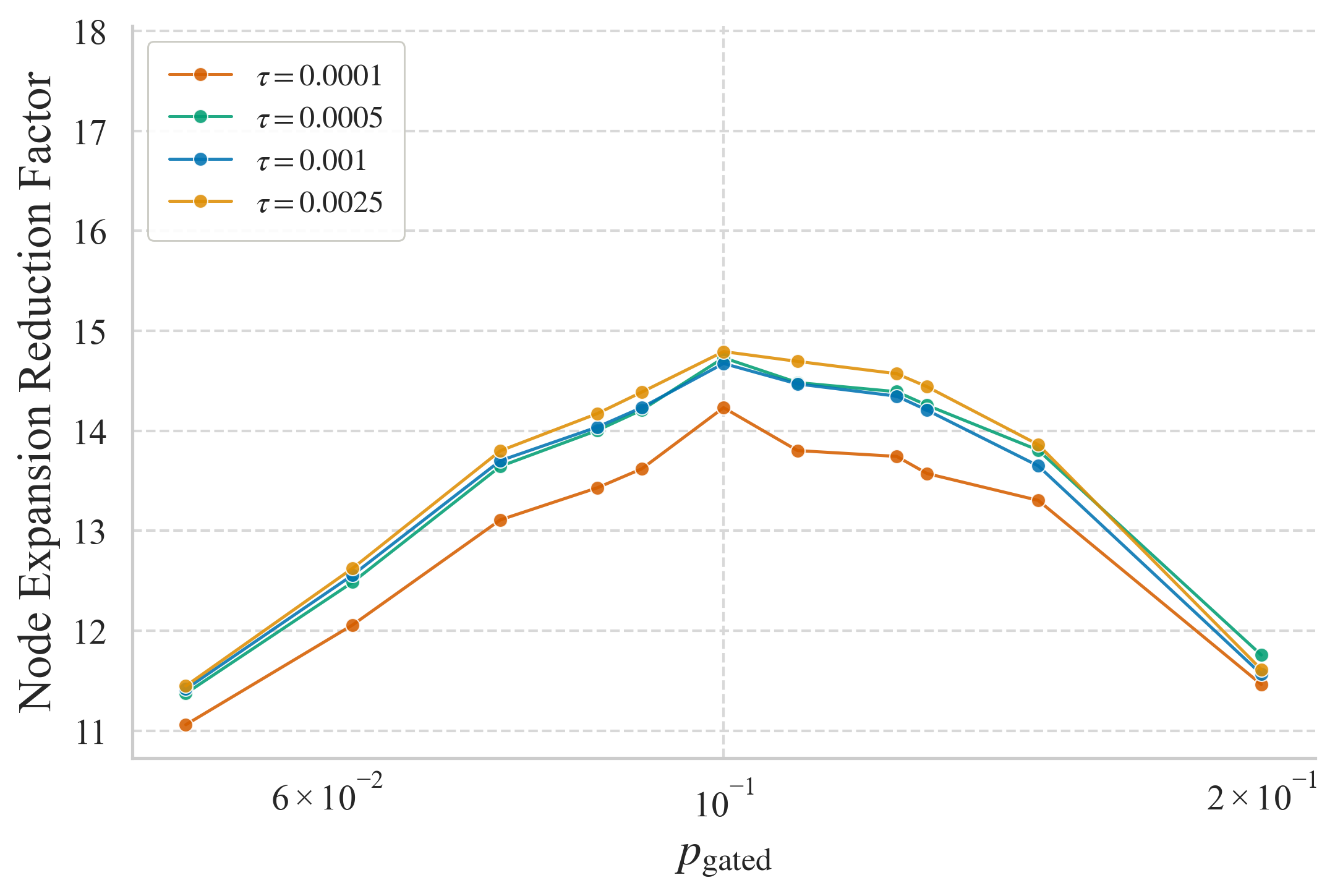}
    \caption{Node expansion reduction factor versus $p_{\mathrm{gated}}$ values for varying $\tau$ at a fixed physical error rate of $p=0.05$. The node expansion reduction factor is defined as the ratio of the number of nodes expanded by the stand-alone Tesseract to those expanded by the BP-Tesseract hybrid decoder.}
    \label{fig:peak0.05}
\end{figure}

When selecting parameter ranges, no constraints were imposed on the value for $\tau$. However, the choice of $p_{\mathrm{gated}}$ is subject to two main constraints. First, it must satisfy $p_{\mathrm{gated}} \in (0, 1/2]$. Values greater than 0.5 would yield negative weights in the cost function (as can be seen from (\ref{EQ: Cost function}) used in Tesseract), causing the search algorithm to behave incorrectly. The second constraint is that the value for $p_{\mathrm{gated}}$ must be greater than the highest $p_{\mathrm{flip},i}$, if not, gated qubits would be assigned a lower cost than reliable qubits. As a result, Tesseract would prioritize modifying these reliable qubits, even though they are likely to be correct. To avoid this, the cost associated with modifying a gated qubit must be higher than that of modifying a reliable qubit. Therefore, $p_{\mathrm{gated}}$ should satisfy 
\begin{equation}
    p_{\mathrm{flip, max}} = \max_i p_{\mathrm{flip, i}},
\end{equation}
where
\begin{equation}
    p_{\mathrm{gated}} \in (p_{\mathrm{flip, max}},1/2].
    \label{EQ: gated interval}
\end{equation}
Figure \ref{fig:peak0.05} shows the reduction factor in the number of node expansions as a function of $p_{\mathrm{gated}}$, where each curve corresponds to a different value for $\tau$ at a fixed error rate of $p=0.05$. The reduction factor is defined as
\begin{equation}
    \text{Node Expansion Reduction Factor}=\frac{N_{\mathrm{A^*}}}{N_{\mathrm{BP-T}}},
\end{equation}
where $N_{\mathrm{A^*}}$ is the number of nodes expanded by stand-alone Tesseract and $N_{\mathrm{BP-T}}$ is the number of nodes expanded by the proposed hybrid BP-Tesseract decoder. The global maximum value of the node expansion reduction factor, 14.79, occurs when $p_{\mathrm{gated}} = 0.1$ and $\tau = 0.0025$. Notably, the optimal value of $p_{\mathrm{gated}}$ appears to be independent of $\tau$, as similar peak values are observed across different $\tau$ settings. When the value of $\tau$ is too high, too many qubits are not gated, which degrades the information given to Tesseract. If $\tau$ is too low, too many qubits are gated, reducing the amount of useful information available and thus increasing the number of node expansions. If $p_{\mathrm{gated}}$ is too low, gated qubits are only slightly penalized, and their cost is not much higher than the reliable qubits, while if the value for $p_{\mathrm{gated}}$ is too high, they may be over-penalized. Since BP reliability is not always accurate, some qubits marked as reliable may still be incorrect. As a result, $p_{\mathrm{gated}}$ must be carefully chosen to avoid both under-penalizing unreliable qubits and over-penalizing potentially correct ones, leading to optimal values for both $\tau$ and $p_{\mathrm{gated}}$. 
\section{Results}
\begin{figure}[tb]
    \centering
    \subfloat[T1 $\mathrm{[[126,12, d<11]]}$.\label{fig:avg_node_T1}]{
        \includegraphics[width=0.472\columnwidth]{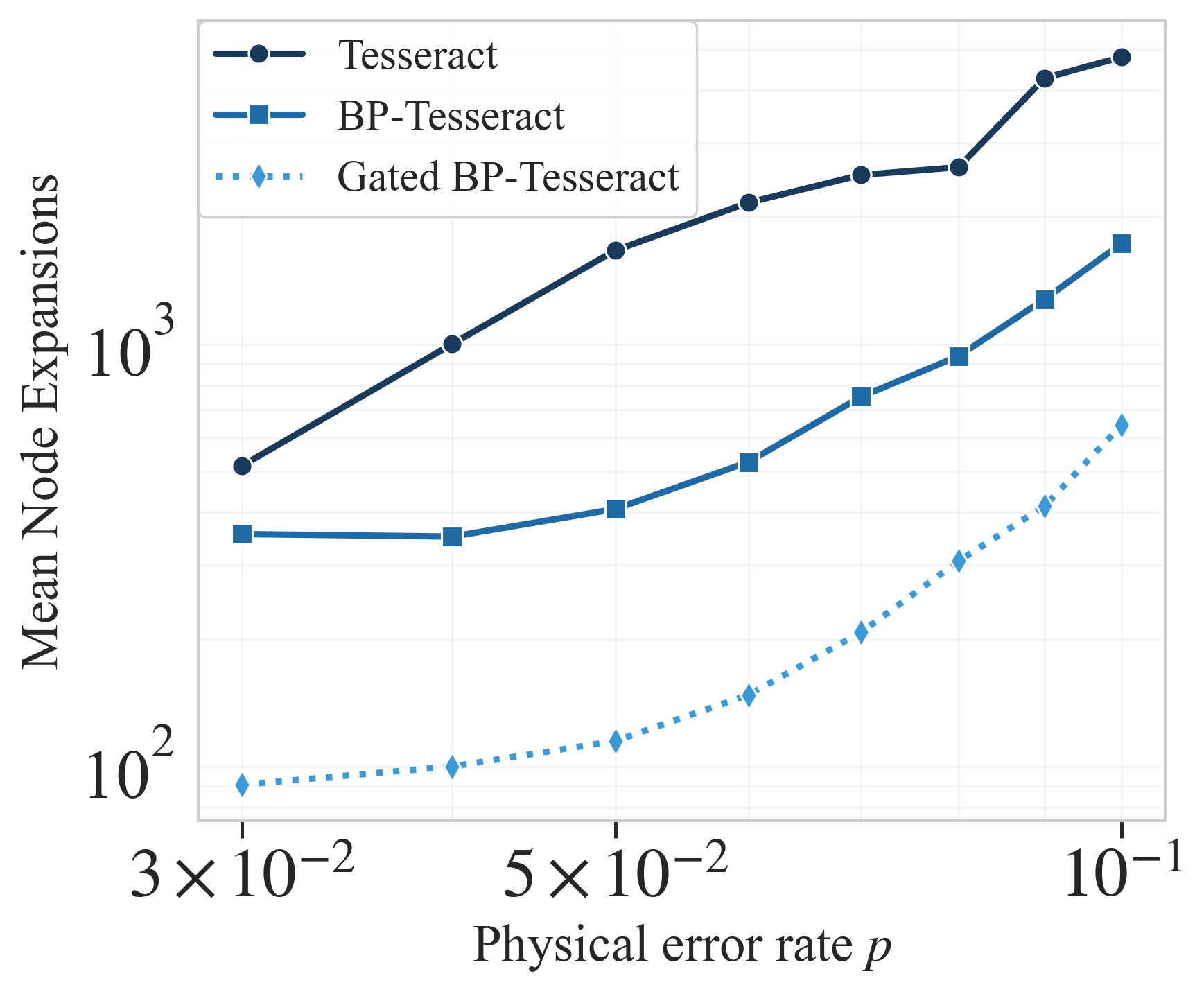}
    }
    \hfill
    \subfloat[BB $\mathrm{[[72, 12, 6]]}$.\label{fig:avg_node_BB}]{
        \includegraphics[width=0.472\columnwidth]{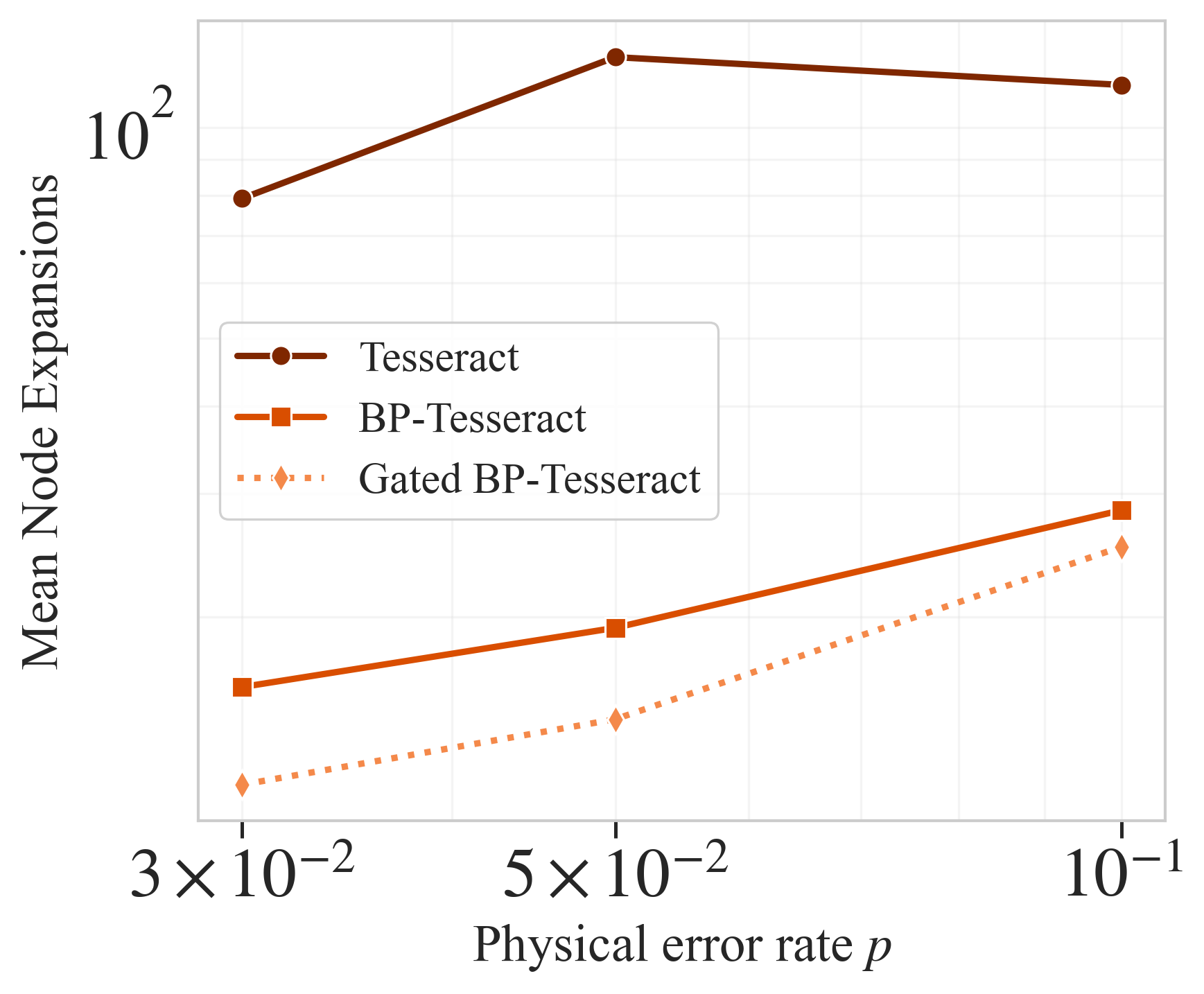}
    }
    \caption{Mean number of node expansions versus physical error rate for the stand-alone Tesseract decoder, the BP-Tesseract decoder, and the gated BP-Tesseract decoder.}
    \label{fig:avg_node}
\end{figure}

The process of finding the peak reduction in the number of expanded nodes, as shown in Figure \ref{fig:peak0.05}, was applied over a range of physical error rates from $p=0.03$ to $p=0.1$ for the $\mathrm{[[126,12, d<11]]}$ T1 code and for the $[[72, 12, 6]]$ bicycle bivariate code, and jointly optimal values of $\tau$ and $p_{\mathrm{gated}}$ were determined. Figure \ref{fig:avg_node} presents a comparison of the average number of node expansions for the stand-alone Tesseract, the hybrid BP-Tesseract, and the gated BP-Tesseract decoders for both the T1 and the bicycle bivariate code.

It can be observed that the gated BP-Tesseract consistently demonstrates a lower average number of node expansions compared to both the stand-alone Tesseract and the BP-Tesseract without probability gating, with reductions reaching nearly $15\times$ for the T1 code. Even at the lowest tested error rate of $p=0.03$, the reduction remains substantial at approximately $5\times$. For the bicycle bivariate code, a reduction of approximately $8.8\times$ is observed at $p=0.05$. The smaller improvement at low physical error rates is expected because fewer Tesseract node expansions are required. Nonetheless, the consistently significant reduction across all tested error rates indicates that the gating mechanism offers substantial computational savings for all physical error rates. The additional improvement over BP-Tesseract confirms the benefit of reliability gating, indicating that passing all BP soft information directly to Tesseract without any filtering is not optimal in non-convergent cases.

\begin{figure}[tb]
    \centering
    \includegraphics[width=\columnwidth]{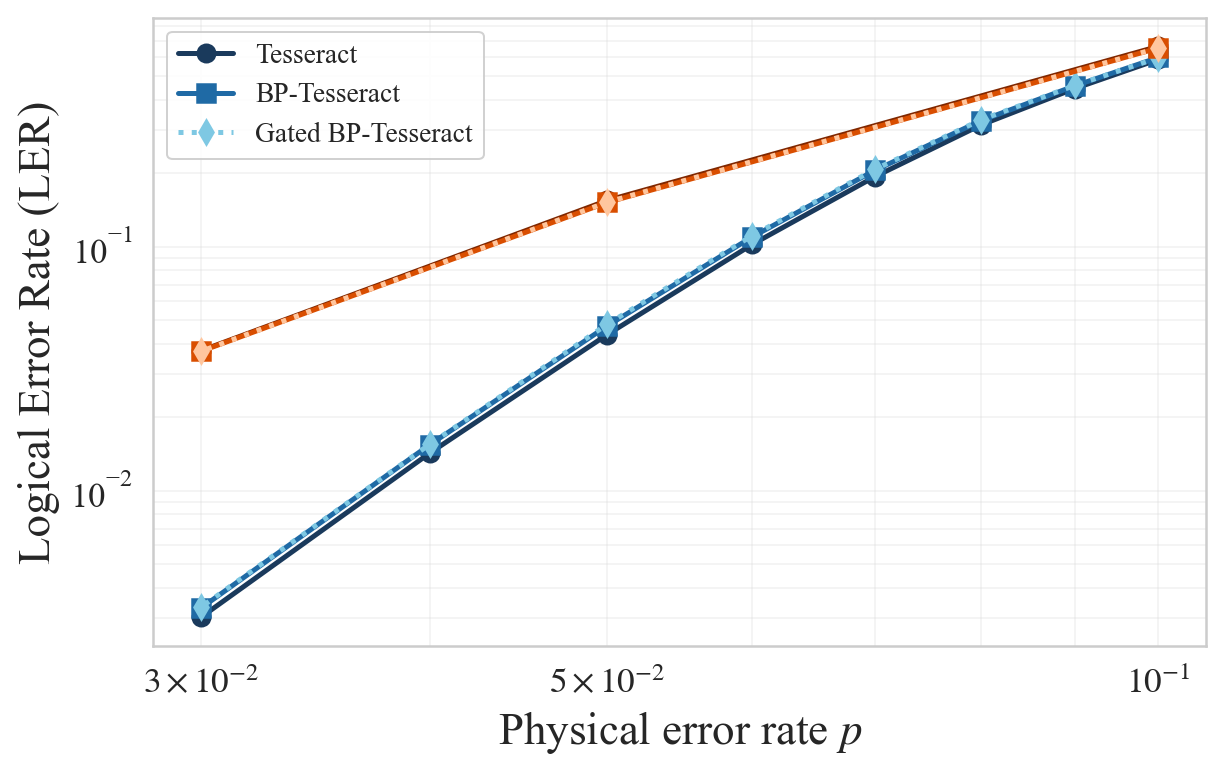}
    \caption{LER performance versus physical error rate for the stand-alone Tesseract decoder, the BP-Tesseract decoder, and the gated BP-Tesseract decoder, for the T1 and BB codes.}
    \label{fig:LER_both}
\end{figure}

While the reduction in node expansions is substantial, it is also important to assess the impact on the logical error rate (LER) performance. Figure \ref{fig:LER_both} presents the LER results for all three decoders for the T1 and bicycle bivariate codes. In both cases, the LER for the proposed decoder is extremely close to that of the original Tesseract. These results indicate that the proposed decoder achieves an effective balance between computational complexity and decoding performance, as it substantially reduces the Tesseract decoding complexity while maintaining LER values close to those of the stand-alone Tesseract. The gating mechanism further enhances this performance by filtering unreliable soft information, leading to additional reductions in complexity, while retaining enough useful information to preserve decoding accuracy.
\section{Conclusion\label{sec:conc} }
In this paper, we have proposed a hybrid BP-Tesseract decoding framework for QLDPC codes that combines the low-latency nature of BP with the strong decoding performance of Tesseract. The BP stage provides soft information and initial probabilistic estimates that change the original decoding task into a simpler problem for Tesseract, i.e., that of correcting the errors made by BP. To further improve the performance of the proposed method, we have also introduced a reliability-based gating mechanism that filters oscillatory and uncertain BP soft information before passing it to the Tesseract stage. Simulation results show that the proposed method achieves a substantial reduction in $\mathrm{A}^*$ search complexity, with up to nearly $15\times$ fewer node expansions, while maintaining LER values comparable to the stand-alone Tesseract.

The results demonstrate that reliability-guided preprocessing can significantly enhance decoding efficiency without substantially compromising accuracy, making this approach suitable for low-latency quantum error correction. Directions for future research include testing on larger QLDPC codes as well as incorporating detector error models in order to demonstrate more fully the scalability of the proposed approach.

\section*{Acknowledgment}

The work of L. Yous and M. F. Flanagan was supported by Science Foundation Ireland under the US-Ireland R\&D Partnership Programme (CoQREATE, Grant Number SFI/21/US-C2C/3750). F.G.H. acknowledges support from the project PID2023-147059OB-I00 funded by MCIU/AEI/ 10.13039/501100011033/ FEDER.  UE. 

\bibliographystyle{IEEEtran}
\bibliography{Bibliography}
\end{document}